\documentclass[fleqn,usenatbib]{mnras}
\usepackage{newtxtext,newtxmath,hyperref}
\usepackage[T1]{fontenc}
\usepackage{graphicx}	% Including figure files
\usepackage{amsmath}	% Advanced maths commands
\usepackage{enumitem}
\usepackage[english]{babel}
\usepackage{gensymb}
\usepackage{hyperref}
\usepackage{comment}
\usepackage{xcolor}
\usepackage{soul}
\definecolor{darkred}{rgb}{0.55, 0.0, 0}

\title[Foreground emission spectral index with LEDA]{Spectral index of the Galactic foreground emission in the $50-87$~MHz range}
\author[M. Spinelli et al.]{M.~Spinelli$^{1,2}$\thanks{E-mail: marta.spinelli@inaf.it}, G.~Bernardi$^{3,4,5}$, H.~Garsden$^{6}$, L.J.~Greenhill$^{6}$, A.~Fialkov$^{7,8}$  J.~Dowell$^{9}$ and D.C.~Price$^{6,10}$ \\
$^{1}$INAF-Osservatorio Astronomico di Trieste, Via G.B. Tiepolo 11, 34143 Trieste, Italy\\
$^{2}$IFPU - Institute for Fundamental Physics of the Universe, Via Beirut 2, 34014 Trieste, Italy\\
$^{3}$INAF-Istituto di Radioastronomia, via Gobetti 101, 40129, Bologna, Italy\\
$^{4}$Department of Physics \& Electronics, Artillery Road, Rhodes University, Grahamstown, South Africa\\
$^{5}$South African Radio Astronomy Observatory, FIR street, Observatory, Cape Town, South Africa\\
$^{6}$Harvard-Smithsonian Center for Astrophysics, 60 Garden Street, Cambridge MA 02138 USA\\
 $^{7}$Kavli Institute for Cosmology, Madingley Road, Cambridge CB3 0HA, UK \\
 $^{8}$Institute of Astronomy, University of Cambridge, Madingley Road, Cambridge CB3 0HA, UK\\
 $^{9}$University of New Mexico, 1919 Lomas Boulevard NE, Albuquerque, NM 87131, USA\\
 $^{10}$International Centre for Radio Astronomy Research, Curtin University, Bentley WA 6102, Australia\\}
\date{Accepted XXX. Received YYY; in original form ZZZ}
\date{\today}
\pubyear{2020}

\begin{document}
\label{firstpage}
\pagerange{\pageref{firstpage}--\pageref{lastpage}}
\maketitle

\begin{abstract} 
Total-power radiometry with individual meter-wave antennas is a potentially effective way to study the Cosmic Dawn ($z\sim20$) through measurement of sky brightness arising from the $21$~cm transition of neutral hydrogen, provided this can be disentangled from much stronger Galactic and extra-galactic foregrounds.  In the process, measured spectra of integrated sky brightness temperature can be used to quantify the foreground emission properties. 
In this work, we analyze  a subset of data from the Large-aperture Experiment to Detect the Dark Age (LEDA) in the range $50-87$~MHz and constrain the foreground spectral index $\beta$ in the northern sky visible from mid-latitudes. We focus on two zenith-directed LEDA radiometers and study how estimates of $\beta$ vary with local sidereal time (LST).  We correct for the effect of gain pattern chromaticity and compare estimated absolute temperatures with simulations. We develop a reference dataset consisting of 14 days of optimal  condition observations. Using this dataset we estimate, for one radiometer, that $\beta$ varies from $-2.55$ at LST~$<6$~h to a steeper $-2.58$ at LST~$\sim13$~h, consistently with sky models and previous southern sky measurements. In the LST~$=13-24$~h range, however, we find that $\beta$ fluctuates between $-2.55$ and $-2.61$ (data scatter $\sim0.01$).  We observe a similar $\beta$ vs. LST trend for  the second radiometer, although with slightly smaller $|\beta|$, in the $-2.46<\beta<-2.43$ range, over $24$~h of LST (data scatter $\sim0.02$). Combining all data gathered during the extended campaign between mid-2018 to mid-2019, and focusing on the LST~$=9-12.5$~h range, we infer good instrument stability and find $-2.56<\beta<-2.50$ with $0.09<\Delta\beta<0.12$.

\end{abstract}


%%%%%%%%%%%%%%%%%%%%%%%%%%%%%%%%%%%%%%%%%%%%%%%%%%%%

\section{Introduction} \label{sec:intro}

The challenging measurement of a signal from the Cosmic Dawn of the Universe finds its most promising observable in the $21$~cm line of the neutral Hydrogen
\citep[e.g.,][]{Furlanetto2006,Pritchard2010}.
At Cosmic Dawn, Ly-$\alpha$ photons, produced by the first stars, couple the excitation temperature of the $21$~cm line 
to the gas kinetic temperature 
through the Wouthuysen-Field effect \citep[WF,][]{Wouthuysen1952, Field1958}, producing a negative contrast
against the Cosmic Microwave Background (CMB) temperature and thus an absorption signal. 
Eventually, the progress of structure formation through gravitational collapse completely couples the spin temperature to the gas temperature. As a consequence of gas heating, most likely by an X-ray background, the gas/spin temperature is then driven well above the CMB temperature  \citep[e.g.,][]{Venkatesan2001,Pritchard2007,Mesinger2013}.
The interplay between Ly-$\alpha$ coupling and X-ray heating is thus expected to create a few hundreds mK absorption features in the {\it global - i.e. sky averaged -} $21$~cm signal, sensitive to the formation of the first luminous structures in the Universe \citep[e.g.,][]{Barkana2005,Furlanetto2006,Fialkov2013,Mirocha2014,Mesinger2016,Mirocha2019}, and to the thermal history of the intergalactic medium \citep{Pritchard2007,Mesinger2013,Fialkov2014}. 

In principle, the global $21$~cm signal can be measured with a single antenna and noise-switched receiver in {\it O}($10^2$) hours \citep[e.g.,][]{Shaver1999,Bernardi2015,Harker2016}.
The Experiment to Detect the Global Epoch-of-Reionization Signatures (EDGES) team has claimed detection of a broad absorption profile peaking at $-520$~mK and centered at $78$~MHz \citep{Bowman2018}.  This is more than a factor two deeper than predicted from theory based on standard physics \citep[e.g. ][]{Cohen2017}. If the EDGES signal is confirmed to be of cosmological origin, it implies that either the temperature of the radio background is higher than that of the  CMB \citep[e.g.][]{Bowman2018} or the neutral gas at $z\sim 17$ is colder than expected  \citep[e.g. due to  interactions with cold dark matter,][]{Barkana2018}. 
The flat line profile is also novel.

Subsequent analysis by other authors have suggested the existence of a  residual unmodeled systematic sinusoidal feature in EDGES data \citep{Hills2018,Singh2019,Bevins2020} that could modify their best fit parameters. Similarly, \citet{Spinelli2019} suggested an unaccounted for contamination from polarized foregrounds. Moreover, using Bayesian evidence-based comparison, \citet{Sims2020} found no strong evidence in favor of models including a global $21$~cm. 
The importance and the implications of the EDGES result need confirmation from another independent experiment to be fully convincing against criticisms. 

\medskip The effort to measure the global $21$~cm signal has been ongoing for several years with variegated instrumental designs. The Large aperture Experiment to detect the Dark Ages \citep[LEDA;][]{Price2018}, whose latest data are analyzed in this work, constrained at $95\%$ confidence level the amplitude ($>-890$~mK) and the $1\sigma$ width ($>6.5$~MHz) of a Gaussian model for the trough \citep{Bernardi2016}. Moreover, the upgraded  Shaped Antenna measurement of the background RAdio Spectrum (SARAS~3) has provided constraints in the $6 < z < 10$ range \citep{Singh2017,Singh2018};  the ``Sonda  Cosmologica  de  las  Islas  para  la Deteccion de Hidrogeno Neutro  \citep[SCI-HI;][]{Voytek2014} reported a $1$~K root mean square (RMS) residual in the range $60-88$~MHz. Other experiments are the  Probing Radio Intensity at high-Z from Marion (PRIZM) experiment \citep{Philip2019}, the Broadband Instrument for Global HydrOgen ReioNisation Signal \citep[BIGHORNS;][]{Sokolowski2015} and the Radio Experiment for the Analysis of Cosmic Hydrogen (REACH)\footnote{\hyperlink{http://www.astro.phy.cam.ac.uk/research/research-projects/reach}{http://www.astro.phy.cam.ac.uk/research/research-projects/reach}}. Moreover, there are proposals
(e.g. the Dark Ages Polarimeter Pathfinder [DAPPER]\footnote{\hyperlink{https://www.colorado.edu/project/dark-ages-polarimeter-pathfinder}{https://www.colorado.edu/project/dark-ages-polarimeter-pathfinder}}) to measure the $21$~cm global signal from the orbit of the moon to avoid terrestrial radio frequency interference, ionospheric corruption and solar radio emissions.

The key challenge to measuring the $21$~cm signal is the subtraction of the bright foregrounds. 
The low-frequency radio  sky  is a superposition of several components including galactic synchrotron and free-free emission, supernova remnants, radio galaxies and absorption from HII regions.
Galactic and extra-galactic synchrotron and free-free emissions have typical sky temperature of thousands of K at 75~MHz, several orders of magnitude stronger than the pristine Cosmic Dawn signal. The proper characterization of the foreground emission thus plays a fundamental role in the $21$~cm global signal analysis.  
Current modeling often uses extrapolations of the all-sky 408 MHz survey of \citet{Haslam1982} to lower frequencies. Other more sophisticated models are available: an improved version of the Global Sky Model \citep[GSM,][]{deOliveira2008,Zheng2017}, including Parkes telescope maps at 150~MHz and 85~MHz \citep{Landecker1970}; the  Global MOdel for the radio Sky Spectrum \citep[GMOSS,][]{GMOSS2017} for the low-frequency radio sky from 22~MHz to 23~GHz. However, the majority of information contained in these models is derived at high frequencies.
Other measurements are available: \citet{Guzman2011} produced a Galactic large-scale synchrotron emission temperature map at 45~MHz covering more than $95\%$ of the sky; the northern hemisphere  has been covered by the Long Wavelength Array (LWA)-1 Low-Frequency Sky
Survey \citep{Dowell2017} at a range of frequencies between 35 and 80~MHz with few degree resolution. 
These maps are not in complete agreement with the GSM at the same frequencies, probably due to an increased contribution of free–free absorption  \citep{Dowell2017}. \citet{Eastwood2018} released low-frequency maps of the full sky visible from the Owens Valley Radio Observatory (OVRO)
between $\sim 36$ and $\sim 73$~MHz. OVRO-LWA maps have $15^\prime$ angular resolution thus complementing the existing full-sky maps at these frequencies, but being exclusively from interferometric observations, they do not represent the globally averaged sky brightness.
 
Global signal experiments, while targeting the Cosmic Dawn, can contribute to the knowledge of the low-frequency radio sky: with their large beams they average the absolute sky brightness temperature over large spatial scales.
The diffuse and continuum foreground sources are known to be spectrally smooth; that is, they should exhibit power-law spectra\footnote{$T(\nu) \propto \nu^{\beta}$} over the low frequency band of interest. Measurements of the Galactic foreground spectral index $\beta$ have been obtained with single dipole antennas.
\citet{Rogers2008} found, for the frequency range $100-200$~MHz, that the spectral index of diffuse emission was $\beta = -2.5\pm 0.1$ at high-Galactic latitudes.
\citet{Patra2015}, using the SARAS experiment, measured the spectral index in the $110-170$~MHz band covering the $23-1$~h local sidereal time (LST) range. They reported a slow variation with LST and a steepening from $-2.3$ to $-2.45$ when observing off the Galactic center.
In \citet{Mozdzen2017} the EDGES team obtained a measurement of the spectral index in the southern hemisphere, in the frequency range $90-190$~MHz.
They found $-2.60 >\beta > -2.62$ in the $0-12$~h LST range, with an increase up to $\beta=-2.50$ at $17.7$~h, when the Galactic center is transiting.
In the $50-100$~MHz range, \citet{Mozdzen2019} found the spectral index to be $-2.59 < \beta < -2.54$ for LST values below $12$~h and a flattening to $-2.46$ when the Galactic center transits.
Moreover, exploiting a lunar occultation technique with the  Murchison Widefield Array telescope, \citet{McKinley2018} measured a spectral index of $-2.64 \pm 0.14$, at the position of the Moon, in the frequency range $72-230$~MHz.

The LEDA experiment covers instead the full northern hemisphere, offering an important complementary measure of the spectral index.
\citet{Price2018}, using LEDA data from $40$ to $83$~MHz, found that the spectral index varies between $-2.28$ to $-2.38$ over the full LST range.

In this work, we present new results on the spectral index of the Galactic foreground emission in the $50-87$~MHz range using the latest LEDA measurements, covering $\sim 140$ nights distributed between mid-2018 to mid-2019.

The paper is organized as follows. In Section~\ref{sec:obs} we describe LEDA observation and data processing, and we assess the quality of our data. In Section~\ref{sec:beam_corr} we present our beam model and correct for the effect of its chromatic response. This is an important step to correctly evaluate the spectral index that was not performed in previous LEDA analysis. 
In Section~\ref{sec:offset} we apply a global temperature scale correction to the data to overcome some systematic effects and to match our spectra with the expected sky temperature.
Our results are presented in Section~\ref{sec:results}, followed by discussion and conclusions in Section~\ref{sec:conclusions}.

\begin{figure}
\includegraphics[width=\columnwidth]{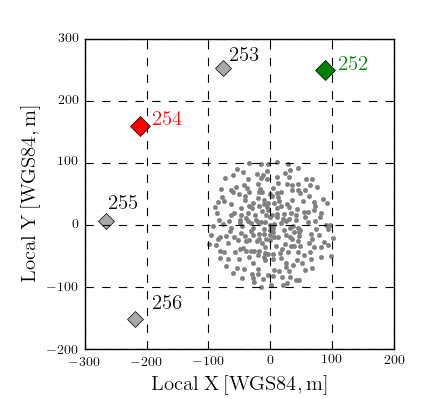}
\caption{OVRO-LWA antenna positions in WGS84 coordinates relative to the center, as built in 2013. Diamonds represent antennas instrumented for radiometry by LEDA.  They are well separated from the core of the array so as to limit mutual coupling effects.} 
We highlight in color the two antennas whose data are used in this analysis.
\label{fig:layout}
\end{figure}

\begin{figure*}
\includegraphics[width=17.5cm]{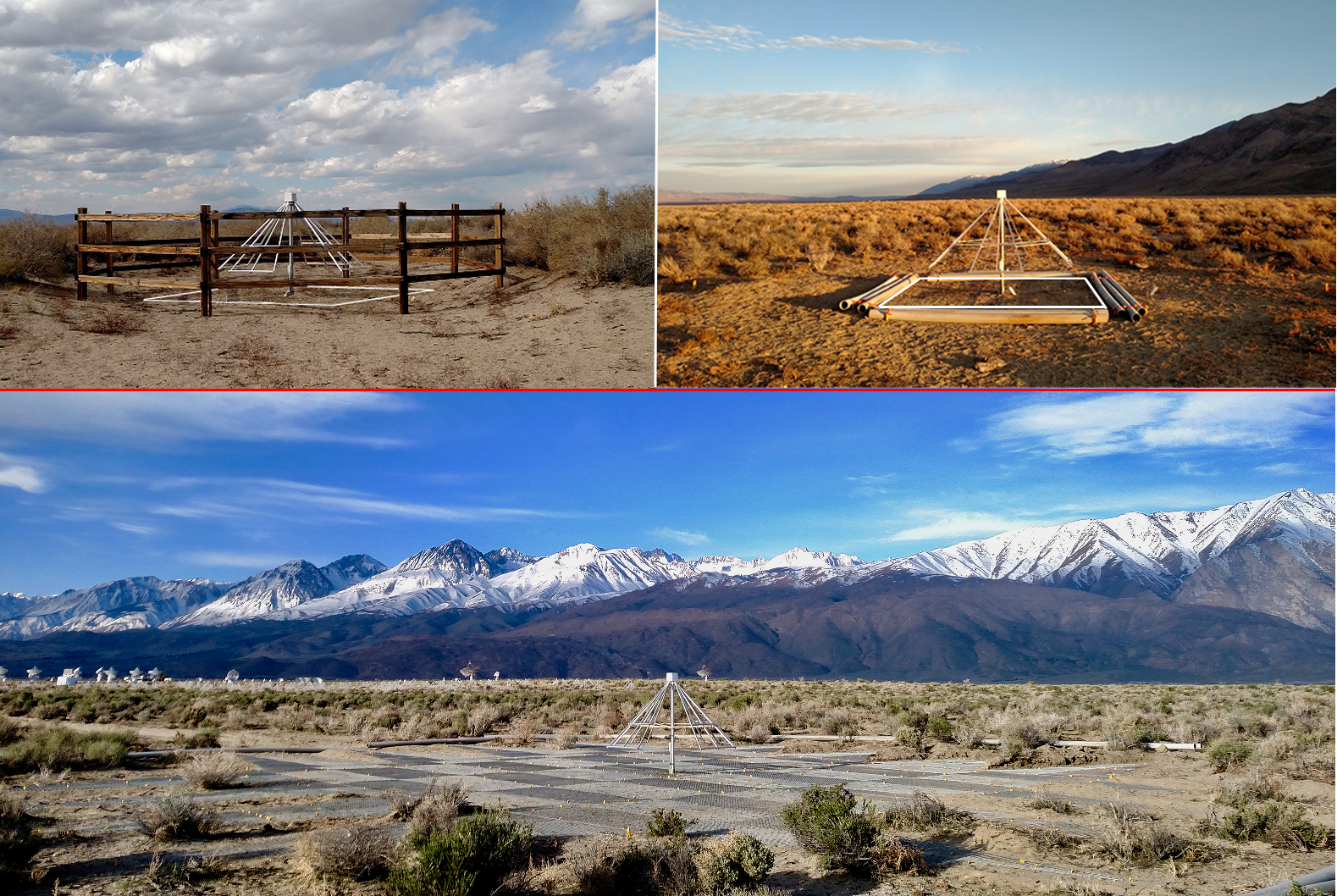}
\caption{LEDA antenna 254, which sits in a slight depression surrounded by $\sim 1$~m high woody vegetation (top left), 252 prior to January 2019 (top right), and 252 after (bottom) completion of the serrated ground screen in early April 2019. Serrations that most strongly affect data were completed in February. The smaller ground screen (top figures) is 3~m~$\times$~3~m (white boundary lines are intended to guide the eye) and flat to 10~cm peak to peak. The larger ground plane (bottom) is 20~m~$\times$~20~m tip to tip and flat to 3~cm peak to peak. The antenna structure is aligned to within $1^\circ$ of true north and the antenna mast stands within $1^\circ$ of vertical. At antenna 252 (top right), PVC pipe was used as a cattle guard in place of wood fencing because water absorbed from rainfall by the wood has the potential to affect antenna gain response.} 
\label{fig:ant_pics}
\end{figure*}

\section{Observations and data processing}\label{sec:obs}

\begin{figure}
\includegraphics[width=\columnwidth]{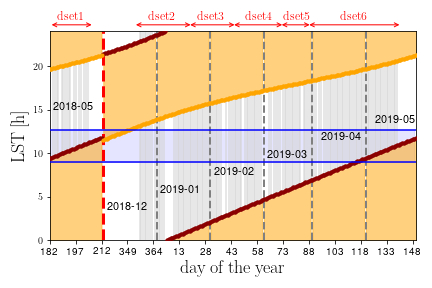}
\caption{The distribution, between May 2018 and May 2019, of the 137 nights analyzed in this work (gray vertical solid lines). The yellow band indicates daylight, defined between sunrise (orange line) and sunset (dark red line). Dashed vertical lines divide the months. Note that there is a gap between May 2018 and December 2018 highlighted with a red,  dashed vertical line. The horizontal blue strip highlights the $\sim 4$~hours LST range that is the main focus of our analysis. We anticipate the partition in six {\it datasets} discussed in Section~\ref{sec:stability} and Figure~\ref{fig:DT_evo} at the top margin of the figure.}
\label{fig:sun}
\end{figure}

Detailed descriptions of LEDA systems can be found in \citet{Kocz2015} and \citet{Price2018}. We  summarize here relevant details  and describe relevant changes to the system  made after data acquisition by \citet{Price2018}.
LEDA dual polarization radiometric receivers were installed in place of LWA receivers on a subset of antennas within the OVRO-LWA ($37^\circ \, 14' \, 23.1998" $~N, $118^\circ \, 16' \, 53.9995"$~W).  
These "LEDA antennas" lie in an arc and are $\sim 170$~m away from others ($\sim 42$~$\lambda$ at 75~MHz), so as to limit mutual coupling effects to $<1\%$ in antenna impedance. The array layout is shown in Figure~\ref{fig:layout}. Each dual-polarization antenna comprises two pairs of triangular dipole arms 1.50~m long, angled downward by $45^\circ$. Antenna terminals contact the receiver $\sim 4.8$~cm above and $\sim 2.2$~cm closer to antenna axis.  The receiver occupies a two sided $\sim 13\times13$~cm circuit board.  A separate daughter board carries those voltage regulators that dissipate far and away the most heat \citep{Price2018}. All components above the dipole arms fit within an uninsulated PVC cubic cover affording  14~cm $\times$ 14~cm $\times$ 11.4~cm of space for hardware. 
The orientation of the east-west dipole axis defines polarization A, and the other polarization B. 
In this analysis we report on results using polarization~A from antennas 252 and 254. 

The ground within 3-4~m of each antenna was devoid of vegetation and flatter than that for other LEDA antennas (i.e., undulations $\ll \pm 10$~cm). Ground screens at each antenna ensured the gain pattern is zenith directed with nulls at the horizon. 

Antenna 254 was equipped with a 3~m $\times$ 3~m ground screen, comprising 12.5~mm gauge welded galvanized wire with a $10.2$~cm spacing.  This was the default for the OVRO LWA station \citep{Paravastu2007, Schmitt2009}. Through the end of January 2019, the same applied to antenna 252.  Afterward, the small screen was replaced by a 20~m $\times$ 20~m ground screen, comprised of 11~mm gauge mesh arranged in a 10~m square with serrations created by four 5~m-long isosceles triangles on each side.  This rework also enabled manual leveling of the ground around antenna 252 to achieve peak-to-peak variation of $\pm 3$~cm. Serrations on the east and west edges were completed on early February, and on the north and south edges on early April (see Figure~\ref{fig:ant_pics}). In general, serrations are used in antenna design to attenuate the otherwise reinforcing effects of reflections due to the impedance mismatch between the ground and the screen along rectangular edges \citep[e.g.,][]{Bowman2018}.

New receiver filters were also installed in order to extend the useful band up to 87.5~MHz.

Observations span a total of 137 days, with 120 days distributed from December 2018 to May 2019, and 17 days in May 2018 as a further control sample for data stability with time. 

In Figure~\ref{fig:sun} we report, for illustrative purposes, the distribution of the 137 days across the year, together with the evolution with LST of sunrise and sunset. Note that the primary LEDA observational window is during night-time hours. The Sun is a potential source of interference, especially during flare events, and RFI environment is known to be more quite at night. Moreover, air traffic and human activity on site is prevalent during daylight hours.

Data were collected with a 24~kHz frequency resolution, over the $30 - 87.5$~MHz range, and radio frequency interference was excised with an iterative thresholding method \citep{Offringa2010,Price2018}. They were calibrated via a three-position switch, which loops between the sky, a cold and a hot load every 5~s, for an actual 30\% time duty cycle on the sky. 
A typical ambient temperature variation may induce up to a 0.2\% power change at the receiver output.
Spectra were further corrected for reflection coefficients measured in the lab \citep{Price2018}. 

As we will discuss further in Section~\ref{sec:offset}, the Galactic plane contributes significantly to the overall system temperature of zenith-directed low-frequency radiometers,  so, for aiming at a detection of the $21$~cm signal, it is better to observe when the brightest portions of the plane are low in the sky. We report in Figure~\ref{fig:sun} the preferred LST range for our analysis, which combines low Galactic plane contribution and night hours across all the observing period.

The consistency of low-frequency radiometer performance depends in part on the environment, the effect of variable soil moisture on complex permittivity 
being prominent. Seasonal dry conditions in 2018 carrying over from summer persisted into the first half of January 2019. Consequently, we anticipate data acquired in December 2018 and January 2019 to be the best. 
This constitutes our {\it reference} dataset.
Unusually heavy winter rains thereafter posed challenges to be addressed in later sections. For the purposes of a more complete analysis of the foreground spectral index, we extend the analysis of December/January data to nearly 24~h (i.e., including sidereal times when bright Galactic emission is high in the sky). 

The inclusion of daytime data allow us to consistently analyze a longer LST range with optimal dry soil conditions.
Daytime spectra do not show noticeable spectral variations that can be attributed to solar activity. We estimate the contribution of the quiet sun at transit in December/January using the solar model from \citet{Benz2009} and find it to be rather smooth in frequency, with a maximum value of $\sim 35$~K at $80$~MHz - i.e. a $\sim 1\%$ contribution to the total sky temperature. According to the Learmonth Solar Radio Spectrograph\footnote{\url{https://www.sws.bom.gov.au/}} monitoring the Sun was essentially quiet during our campaign.

\begin{comment}
In Figure~\ref{fig:waterfall} we present, for illustrative purposes, 24~h observation from December 2018. We can appreciate variations in the foreground amplitude as a function of LST.
\end{comment}

\subsection{Statistics of data integration and noise levels}\label{sec:errors}

At frequencies below $\sim 50$~MHz, where the wavelength is twice the size of the default $3\times3$~m ground screen configuration, spectra change noticeably over time scales of a few days, likely due to the instrument sensitivity to environmental effects. We therefore discarded data below 50~MHz, retaining 1540 of the original 2400 channels. 

For a single frequency channel, the measured temperature varies over time and includes random Gaussian noise, $T(t) + n$. Assuming the noise in each channel to be distributed as a Gaussian with mean $\mu_n = 0$ and standard deviation $\sigma_n$, it is possible to estimate the noise variance  by removing $T(t)$.
We de-trend the signal using a mean filter with a sliding-window of $\sim 4$~minute width (16 data points) to this aim.
This procedure does not inject correlation into the data, but
the uncertainties will be partially correlated over the 4 minute interval. 
Once $T(t)$ is removed, we verify that the extracted noise is Gaussian distributed and estimate the variance $\sigma_n$ of the distribution. In the range of frequencies we are interested in for our analysis, the $\sigma_n$ values per channel are around tens of Kelvin. 

We averaged the signal over 1~MHz-wide channels, using the variance computed in each channel as a  weight. The typical noise in a 1~MHz channel is a few Kelvin at 75~MHz for a 5~s integration time.
These 1~MHz spectra are the ones used in our analysis.

\subsection{Stability and consistence}\label{sec:stability}
\begin{figure}
\includegraphics[width=\columnwidth]{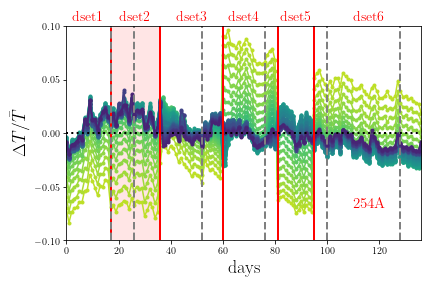}
\includegraphics[width=\columnwidth]{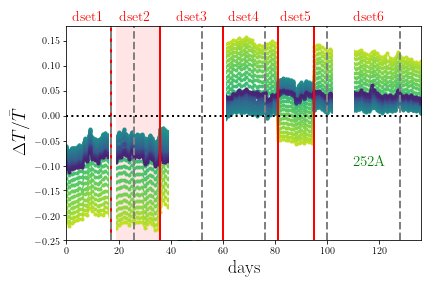}
\caption{Variations of the measured sky temperature $T_{\rm d}(\nu)$, for 1~h integration time around LST $=12$~h, as a function of observing day (d), for antenna 254A (upper panel) and 252A (lower panel). We plot the relative difference between $T_{\rm d}(\nu)$ and its mean value $\bar{T}$(equation~\ref{eq:Tbar}). Different colors indicate different frequencies: lower (higher) frequencies are shown in lighter greens (darker blues). 
Dashed vertical lines denote the separation between months. May 2018 appears at the beginning of the x axis (note the dashed line superimposing with the first red vertical line), followed by December 2018 up to May 2019. Note that observations do not span all days of every month, therefore the number of days between two consecutive vertical dashed lines is smaller than 30.
The relative temperature shows time discontinuities common to both antennas, more pronounced at low frequencies. These are highlighted in the plot with red vertical lines and define the six different 
{\it datasets}. December 2018 and January 2019 data (dset2), marked as a red shaded region, constitute our {\it reference} dataset as they are optimal in term of soil condition.
Note that, despite a less prominent discontinuity, we keep May 2018 separated from December 2018.}
\label{fig:DT_evo}
\end{figure}

The observed temperature varies as a function of LST as the foreground sky changes as observed by a fixed antenna, however, it is ideally the same if measured at the same LST over different days. Figure~\ref{fig:DT_evo} shows how the observed temperature varies over the course of our observations.
For each frequency channel, we plot the relative variation of the observed temperature, calculated against the time average over N days: \begin{equation}\label{eq:Tbar}
\bar{T}(\nu)=\frac{1}{N}\sum_{{\rm d}=1}^{\rm N} T_{{\rm d}}(\nu), \;{\rm with} \; {\rm N}=137.
\end{equation}
Temporal variations are frequency-dependent in both antennas, with larger effects at low frequencies.
We mark them in the Figure with red vertical lines and define in this way six different {\it datasets}, where temperature remains fairly constant with time.
Despite a less sharp change of trend, we kept the data of May 2018 separated from the ones of December 2018.
We recall that the December/January data are our {\it  reference} dataset due to the seasonal optimal conditions for humidity and rainfall.
The time discontinuities appear more evident in antenna 252A although contained at few $\%$ level, reaching $10-20\%$ only at the lowest frequencies ($\nu < 60 $~MHz). 
In this example we consider 1~h integration time around LST $=12$~h as a representative trend visible at other LST values too.

\begin{figure}
\includegraphics[width=\columnwidth]{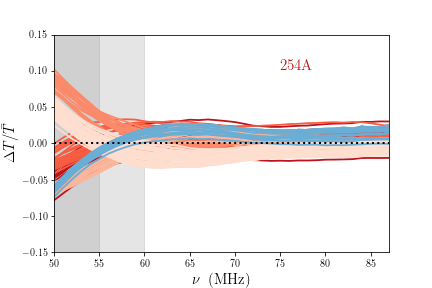}
\includegraphics[width=\columnwidth]{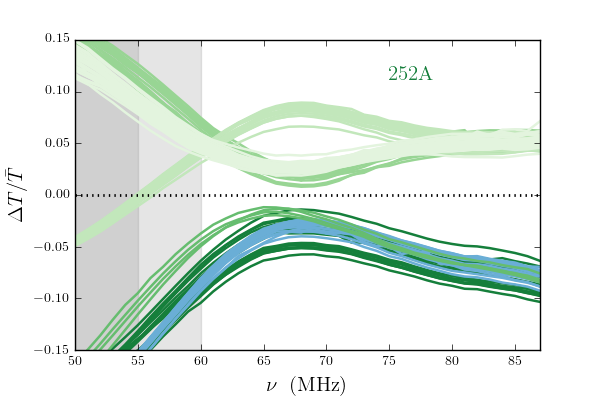}
\caption{Time variation of the sky spectra measured by antenna 254A (upper panel) and 252A (lower panel), plotted as the relative difference with respect the mean value $\bar{T}$ - each color is a different {\it dataset} as defined in Figure~\ref{fig:DT_evo}. Chronologically earlier (later) datasets are shown in darker (lighter) colors (red for antenna 254A and green for antenna 252A). The dset2 {\it reference} dataset appears in blue in both plots in order to facilitate its identification. Gray shaded areas identify the lowest frequencies (dark gray $\nu < 55$~MHz, light gray $\nu < 60$~MHz). }
\label{fig:DT_nu}
\end{figure}

We then study the variation over time of the spectra measured by the two antennas. We show in Figure~\ref{fig:DT_nu} the relative difference with respect to the mean value $\bar{T}$, computed by averaging all measured spectra. We plot the six {\it datasets} defined above with different colors. The 254A data show good stability; for frequencies higher than 60~MHz, within-datasets and between-datasets differences are at most $5\%$
and do not evolve much with frequency. Below 60~MHz, the different trends highlighted in Figure~\ref{fig:DT_evo} are again clearly visible, leading to differences up to 10$\%$. 
Antenna 252A has good consistency within each dataset, while differences across datasets can be as high as 30$\%$. These latter are due both to temporal offsets as well as low frequency variations, visible in Figure~\ref{fig:DT_evo} also. Differences are smaller at higher frequencies. We note that antenna 252A has a particularly marked discontinuity between January and February, corresponding to maintenance activities.
Further details on the amount of measurements available per LST slots in our analysis can be found in Appendix~\ref{app:data}.

\section{Beam Chromaticity}\label{sec:beam}
The spectral smoothness of the foreground emission is the key property that should allow the retrieval of the $21$~cm signal.
Regardless of the parametrization used for the foregrounds, a frequency-dependent response of the antenna could corrupt their intrinsic spectra and thus prevent the extraction of the global signal \citep[e.g.][]{Bernardi2015,Mozdzen2016, Tauscher2020, Anstey2020}.
This beam chromaticity effect, however, can be (at least) partially compensated for by assuming an antenna beam and a foreground model \citep[e.g.][]{Mozdzen2017,Mozdzen2019}. 

\subsection{Beam model}\label{sec:beam_model}

We use the NEC-4 package\footnote{\url{https://ipo.llnl.gov/technologies/software/nec-v50-numerical-electromagnetic-code}} that employs a method of moments to simulate an isolated dipole with a 3~m~$\times$~3~m mesh ground screen, a $200$~$\Omega$ load seen by the antenna at its interface with the receiver and a constant soil conductivity of 0.001~S~m$^{-1}$. The conductivity value is somewhat more representative of the December/January drier period. 

\begin{figure}
\includegraphics[width=\columnwidth]{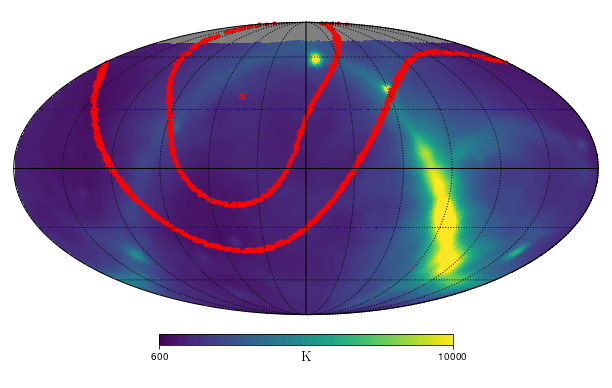}
\caption{Red contours show -3dB and -10dB projection, in celestial coordinates, of the E-W oriented LEDA antenna model at 75~MHz, for LST~$=3$. The Guzman 45~MHz map scaled at 75~MHz with a constant spectral index of $-2.5$ is shown in the background for illustration purposes. The gray area shows the missing data at $\rho>65^{\degree}$  As the sky drifts the antenna is seeing different structures at the different LSTs.}
\label{fig:antenna_hp}
\end{figure}

Simulations were used to constrain an analytical beam model \citep{Dowell2011,Ellingson2013,Bernardi2015}:
\begin{equation}
B(\nu,\mathbf{\hat{n}})=\sqrt{[p_E(\theta,\nu) \cos{\phi}]^2 + [p_H(\theta,\nu) \sin{\phi}]^2},
\notag
\end{equation}
where $\mathbf{\hat{n}}=(\theta,\phi)$, $E$ and $H$ are the two orthogonal polarizations of the dipole
and
\begin{equation}\label{eq:beam}
p_i(\nu,\theta)=\left[ 1 - \left( \frac{\theta}{\pi/2} \right)^{\alpha_i(\nu)} \right](\cos{\theta})^{\beta_i(\nu)} + \gamma_i(\nu) \left(\frac{\theta}{\pi/2} \right)(\cos\theta)^{\delta_i(\nu)}
\end{equation}
with $i=E, H$. The value of the coefficients $[\alpha_i,\beta_i,\gamma_i, \delta_i]$ are fitted with an 11-th order polynomial.
A projection of the resulting beam in celestial coordinates can be found in Figure~\ref{fig:antenna_hp}.

\begin{figure}
\includegraphics[width=\columnwidth]{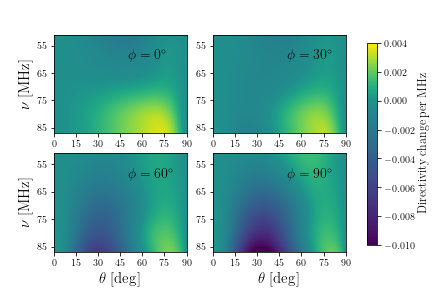}
\caption{Beam directivity change with respect to frequency (considering 1 MHz steps) at $\phi=0^\circ$ (E-plane) and $90^\circ$ (H-plane). For completeness we show also the values for $\phi=30^\circ$ and $60^\circ$.
The horizon response is at $\theta \sim 90^\circ$.}
\label{fig:beam_dir}
\end{figure}

In order to show the shape of the beam and its frequency structure we report in Figure~\ref{fig:beam_dir} the beam derivative with respect to frequency versus zenith angle $\theta$, for different values of the azimut angle $\phi$,  as in \citet{Mozdzen2016}. 
The absolute value of the derivative increases with frequency as expected from increased chromaticity and the more complex antenna pattern in the direction perpendicular to the main axis. Nevertheless, the pattern does not show rapid variations, ensuring beam smoothness.

\subsection{Foreground model}\label{sec:sky_model}
We first consider the Haslam $408$~MHz sky map $T_{{\rm H}}(\mathbf{\hat{n}})$ \citep{Haslam1982} to model foregrounds and scale it to any frequency $\nu$ with a constant spectral index across the entire sky, after subtracting the CMB temperature ($T_{{\rm cmb}}=2.725$~K):
\begin{equation}\label{eq:T_H}
    T^{{\rm H}}_{{\rm sky}}(\nu,\mathbf{\hat{n}})=[T_{{\rm H}}(\mathbf{\hat{n}})-T_{{\rm cmb}}]\left( \frac{\nu}{408}\right)^{-2.5}+T_{{\rm cmb}}.
\end{equation}
A second foreground model is derived from the $45$~MHz Guzman sky map $T_{{\rm G}}(\mathbf{\hat{n}})$ \citep{Guzman2011}.
As this map misses data at declination $\delta>65^\degree$, we inpaint this empty region using $T^{{\rm H}}_{{\rm sky}}(\nu_{45},\mathbf{\hat{n}})$.
We call the resulting map $\tilde{T}_{{\rm G}}(\mathbf{\hat{n}})$.
We can compute from the Haslam and (inpainted) Guzman maps a more accurate "wide-band" spectral index $\beta_{\rm GH}(\mathbf{\hat{n}})$: 
\begin{equation}\label{eq:betaGHn}
    \beta_{\rm GH}(\mathbf{\hat{n}})=\ln{\frac{\tilde{T}^{G}(\mathbf{\hat{n}})}{T^{\rm H}(\mathbf{\hat{n}})}} \, \left( \ln{\frac{45}{408}} \right)^{-1}.
\end{equation}
The final sky model is constructed similarly to equation~\ref{eq:T_H}\footnote{After this work was already submitted, \citet{Monsalve2020}  published a correction for the zero-level of the Guzman map of $-160$~K, with an uncertainty of $78$~K (at $2\sigma$ confidence level). This correction will be taken into account in future work.}:
\begin{equation}\label{eq:T_GH}
    T^{\rm G}_{{\rm sky}}(\nu,\mathbf{\hat{n}})=[\tilde{T}_{{\rm G}}(\mathbf{\hat{n}})-T_{{\rm cmb}}]\left( \frac{\nu}{45}\right)^{\beta_{\rm GH}(\mathbf{\hat{n}})}+T_{{\rm cmb}}.
\end{equation}
\subsection{Beam chromaticity correction}\label{sec:beam_corr}
The beam chromaticity correction can be computed as:
\begin{equation}\label{eq:beam_chromaticity}
B_{{\rm c}}(\nu,{\rm LST})=\frac{\int_{\Omega} T_{{\rm sky}}(\nu_0,{\rm LST},\mathbf{\hat{n}'}) B(\nu,\mathbf{\hat{n}}') d\mathbf{\hat{n}}'}{\int_{\Omega} T_{{\rm sky}}(\nu_0,{\rm LST},\mathbf{\hat{n}}') B(\nu_0,\mathbf{\hat{n}}') d\mathbf{\hat{n}}'}.
\end{equation}
Note that $T_{{\rm sky}}$ is a function of LST since the sky drifts with time over the antenna.
We choose $\nu_0=75$~MHz for the reference frequency in order to compare our results with \citet{Mozdzen2019}. Note that this is a reasonable choice for us regardless, since it is a quite central frequency for the range of our data.
We compute the beam chromaticity using our beam model (Section~\ref{sec:beam_model}) and both the sky model of equation~\ref{eq:T_H} and equation~\ref{eq:T_GH}. We find very similar results, differing almost everywhere  by less than a few tenths of a percent.
The disagreement is slightly more pronounced closer to the Galactic center direction, but never higher than 1\%.  
To test the impact of limitations in our beam model, we instead perturb the coefficients of the beam directivity $p = [\alpha_i,\beta_i,\gamma_i, \delta_i]$ (see equation~\ref{eq:beam}) by drawing values from a Gaussian distribution with standard deviation corresponding to $\sigma_p = 1\%$, $5\%$ and $10\%$ of the original parameter value \citep[similar to what was done in][]{Bernardi2015}. We construct 100 beam models by varying simultaneously all four $p$ parameters for each $\sigma_p$ value and compute the chromaticity corrections. We find variations always smaller than 1\% even in the worst case scenario of $\sigma_p = 10\%$.
We anticipate here that they translate in an error on the spectral index estimation that is smaller than the typical error bar due to the variability of our spectra.

\begin{figure}
\includegraphics[width=\columnwidth]{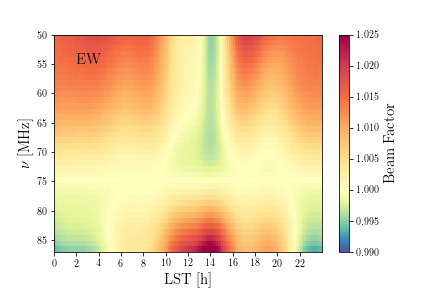}
\caption{Beam chromaticity correction for a simulated E-W oriented LEDA antenna (pol A) as a function of frequency and LST. See text for details.}
\label{fig:beam_chromaticity}
\end{figure}

In Figure~\ref{fig:beam_chromaticity} we present results using our reference beam model and the Haslam sky model to show consistency with Figure 3 of \citet{Mozdzen2019}. 

For our final analysis we compute the chromaticity correction using $T^{\rm G}_{{\rm sky}}$ and for each LST and frequency we use $B_{{\rm c}}(\nu,{\rm LST})$ to correct our spectra:
\begin{equation}
    T_{\rm c}^{\rm obs}(\nu,{\rm LST}) \equiv T^{\rm obs}(\nu,{\rm LST}) / B_{{\rm c}}(\nu,{\rm LST}).
\end{equation}

\section{Absolute temperature scale}\label{sec:offset}

\begin{figure}
\includegraphics[width=\columnwidth]{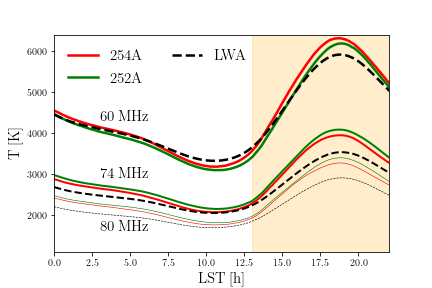}
\caption{Comparison between the {\it reference} dataset mean sky temperature as a function of LST (antenna 254A in red and 252A green) with the average values of the LWA maps from \citet{Dowell2017} weighted by antenna response (equation~\ref{eq:T_LEDA}) at 60, 74 and 80 MHz (black dashed lines with decreasing thickness). For illustrative purposes, the light yellow shaded area indicates the daylight time for the LEDA observations, starting approximately from sunrise.}
\label{fig:T_vs_LWA}
\end{figure}

\begin{figure}
\includegraphics[width=\columnwidth]{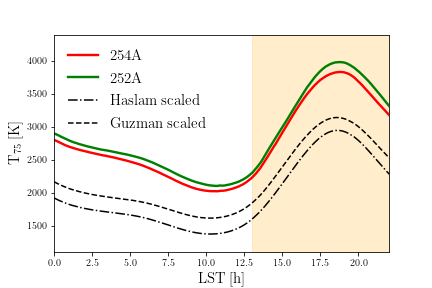}
\caption{The averaged sky temperature measured by the antennas 254A (red) and 252A (green) at $75$~MHz as a function of LST for the {\it reference} dataset.For illustrative purposes, the light yellow shaded area indicates the daylight time for the LEDA observations, starting approximately from sunrise. In the Figure we also report the expected behavior of the mean beam weighted sky temperature (equation~\ref{eq:T_LEDA}) at $75$~MHz from other existing measurements: the black dot-dashed line is obtained with the extrapolated Haslam sky model $T^{\rm H}_{{\rm sky}}$ (equation~\ref{eq:T_H}) and the black dashed from the extrapolated Guzman sky model $T^{\rm G}_{{\rm sky}}$ (equation~\ref{eq:T_GH}).}
\label{fig:T75_vs_model}
\end{figure}

We investigate the behavior of the chromaticity corrected spectra $T_{\rm c}^{\rm obs}(\nu,{\rm LST})$ as a function of LST,
over the 24 hour range (Figure~\ref{fig:T_vs_LWA}, \ref{fig:T75_vs_model}).  We focus on  the {\it reference} dataset defined in Section~\ref{sec:obs} and Figure~\ref{fig:DT_evo}\footnote{As discussed in Appendix~\ref{app:data}, when extended to the 24 hour range, few days of the {\it reference} dataset have been discarded.}.
The measured sky temperature is consistent for the two antennas: it has a minimum in the LST range $10-12$~h and reaches its peak around LST $\sim 18$~h, when the Galactic center is closer to the visible sky.

We compare our results against existing foreground models and measurements by evaluating a simulated antenna temperature:
\begin{equation}\label{eq:T_LEDA}
T^{\rm ant}(\nu,{\rm LST})= \frac{\int_{\Omega} T_{{\rm sky}}(\nu,{\rm LST},\mathbf{\hat{n}})B(\nu_0,\mathbf{\hat{n}}) d\mathbf{\hat{n}}}{\int_{\Omega} B(\nu_0,\mathbf{\hat{n}}) d\mathbf{\hat{n}}},
\end{equation}
where $T_{{\rm sky}}$ are, alternatively LWA1 Low Frequency Sky Survey results \citep{Dowell2017}, the Haslam sky model, $T^{\rm H}_{{\rm sky}}$ (equation~\ref{eq:T_H}), or the Guzman sky model, $T^{\rm G}_{{\rm sky}}$ (equation~\ref{eq:T_GH}). 
Note that we use the beam pattern at the reference frequency $\nu_0=75$~MHz in order to avoid the introduction of chromaticity effects.

\citet{Dowell2017} have corrected their LWA interferometric observations with the sky-averaged temperature measured in December 2014 by the LEDA prototype system, LEDA-64 New Mexico deployment \citep{Taylor2012,Schinzel2018}.
In Figure~\ref{fig:T_vs_LWA}, we report the 60, 74 and 80~MHz frequencies and compare with our chromaticity corrected measurements of the sky temperature obtained with antenna 254A and antenna 252A.
Their trend with LST is very similar and absolute values are consistent within  $\sim15$\%.

In Figure~\ref{fig:T75_vs_model} we compare our measured temperature at 75~MHz for both antennas against the simulated temperature $T^{\rm ant}(75\;{\rm MHz},{\rm LST})$ (equation~\ref{eq:T_LEDA}), where we use either the Haslam sky model $T^{{\rm H}}_{\rm sky}$  or the Guzman sky model $T^{{\rm G}}_{\rm sky}$.
Predictions and our measurements have a comparable behavior with LST, however, they are discrepant in the absolute temperature scale. Although future analysis will be dedicated to investigating the precision of the absolute calibration, in this work we correct our absolute temperature scale to match the extrapolation from the 45~MHz map. We make the conservative choice of  evaluating $T_{\rm off}$ at LST$=12$~h, where both the sky temperature and the offset have a minimum. Moreover, the LST$=12$~h bin is the only one covered across the whole observing campaign and, therefore, the most appropriate choice for the correction.

\begin{figure}
\includegraphics[width=\columnwidth]{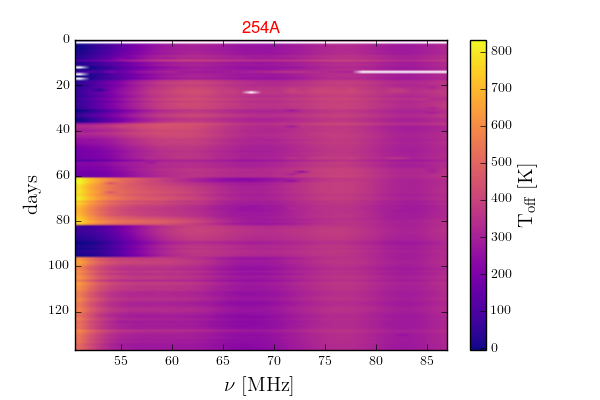}\\
\includegraphics[width=\columnwidth]{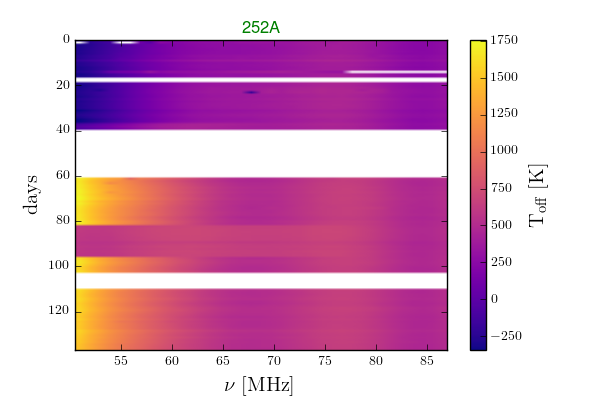}
\caption{Offset temperature $T_{\rm off}(\nu)$ at LST~$=12$~h (see equation~\ref{eq:Toff}) for antenna 254A (upper panel) and 252A (lower panel) as a function of frequency and observing days. Note that the color bars have different extents for the two antennas. 
The difference among {\it datasets} is particularly evident at low frequency.}
\label{fig:Toff}
\end{figure}

\medskip
Figure~\ref{fig:Toff} shows $T_{\rm off} (\nu)$, i.e the difference between all our spectra (corrected for the chromaticity factor) and $T^G_{\rm sky}$ at LST~$=12$~h: 
\begin{equation}\label{eq:Toff}
    T_{\rm off} (\nu) \equiv T_c^{\rm obs}(\nu, {\rm LST}=12{\rm h}) - T^G_{\rm sky}(\nu, {\rm LST}=12{\rm h}).
\end{equation}
The offset is relatively constant with frequency and observing day. The strongest differences at low frequencies highlight the different {\it datasets} defined in Figure~\ref{fig:DT_evo}.
An oscillatory pattern in $T_{\rm off} (\nu)$ with a period of approximately 20~MHz,  common to both antennas, and visible across all the observations,  can be tentatively inferred from Figure~\ref{fig:Toff}, indicating a possible systematic effect still present in the data.

Offset values for antenna 254A closely follow a Gaussian distribution for each {\it dataset}, and the standard deviation becomes smaller if we restrict the range to $60 - 87$~MHz. 

\begin{figure}
\includegraphics[width=\columnwidth]{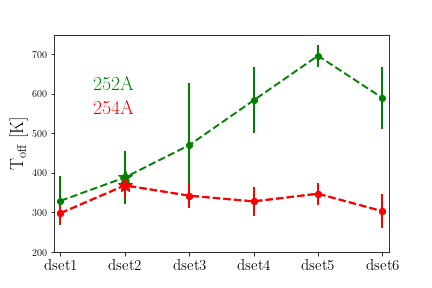}
\caption{Mean values and 1$\sigma$ errors obtained from a Gaussian fit of the $T_{\rm off}$ distributions (Figure~\ref{fig:Toff}). The {\it reference} dataset is marked with a star.}
\label{fig:meanToff}
\end{figure}

The offset values for antenna 252A span a larger range, as can be seen also in Figure~\ref{fig:Toff}, and their distribution  deviates more significantly from a Gaussian profile.
We retain for both antennas only $T_{\rm off}(\nu)$ in the $60 - 87$~MHz range and we fit the various distributions with a Gaussian profile to obtain a mean $\bar{T}_{{\rm off}}$ and standard deviation value for each {\it dataset}.
The best-fit mean and standard deviation are reported in Figure~\ref{fig:meanToff} for both antennas. The results for antenna 254A are quite stable along the duration of the observations while 252A shows a change of trend, consistently with what has been seen already in 
Figure~\ref{fig:DT_evo} and \ref{fig:DT_nu}. 
We correct our spectra by subtracting the mean value $\bar{T}_{{\rm off}}$
for each {\it dataset}.

\begin{figure*}
\includegraphics[width=\columnwidth]{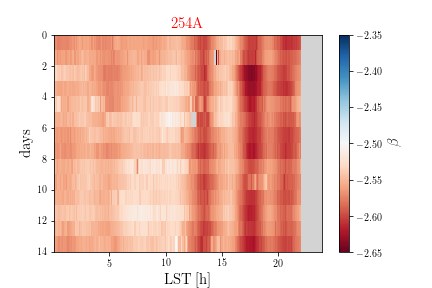}
\includegraphics[width=\columnwidth]{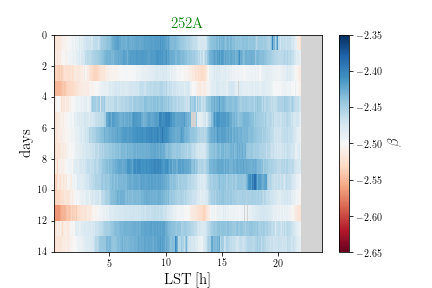}\\
\includegraphics[width=\columnwidth]{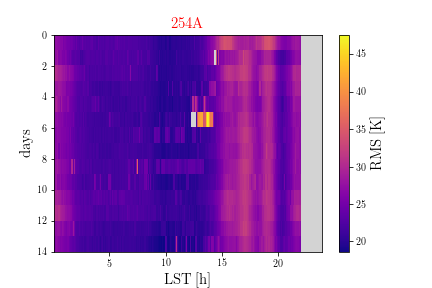}
\includegraphics[width=\columnwidth]{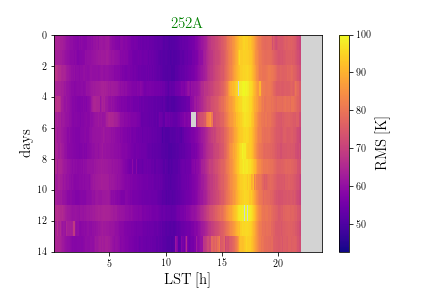}
\caption{{\it Upper panels:} Best fit value for the $\beta$ parameter of equation~\ref{eq:model_beta} for our {\it reference} dataset as a function of LST for the antenna 254A (left) and 252A (right) Note that the colorbar is the same for the two antennas to help the comparison. {\it Lower panels:} The RMS of the residual spectra for antenna 254A (left) and 252A (right).}
\label{fig:waterfall_beta_dec_bc_Toff}
\end{figure*}

\section{Spectral Index Results}\label{sec:results}
\label{sec:spectral_index}
We now analyze our observations to characterize the spectral index of the foreground emission.  
We focus on our {\it reference} dataset 
in Section~\ref{sec:beta_reference}.  We recall that for this dataset the observing conditions are optimal in terms of soil and rainfall. We study the spectral index averaging the spectra (measured in 14 selected days) in LST bins of 5 minutes and across an LST range of $\sim 24$~h. 
We concentrate instead on the $9-12.5$~h LST range, that corresponds to the minimum of foreground contamination, when studying the all six {\it datasets} in Section~\ref{sec:beta_6datasets}. Note that this LST range is the central interval of interest for further studies aimed at extracting a global signal constraint/upper limit. 
We model the data with a simple power law. Following \citet{Mozdzen2019} we write
\begin{equation}\label{eq:model_beta}
    T_{\rm m}(\nu ; \beta, {\rm T}_{75})={\rm T}_{75} \left( \frac{\nu}{\nu_{75}} \right)^{\beta}+T_{\rm cmb},
\end{equation}
choosing $75$~MHz as a reference frequency.
We perform a non-linear least squares minimization to constrain the model parameters for each LST bin, using
\begin{equation}\label{eq:chi}
    S=\sum_i^{N_f}\frac{[({\rm T}^{\rm obs}_c(\nu_i)-\bar{T}_{\rm off})-{\rm T}_{{\rm m}}(\nu_i)]^2}{\sigma_i^2}
\end{equation}
where $N_f$ is the total number of $1$~MHz frequency bins, and the errors $\sigma_i$ correspond to the noise variance estimates discussed in Section~\ref{sec:errors}.
We recall that $\bar{T}_{\rm off}$ varies across {\rm dataset}s (Figure~\ref{fig:meanToff}). In appendix~\ref{app:stability}, we analyze the effect on the fit of the dispersion around this mean.

\subsection{Reference dataset results}
\label{sec:beta_reference}

Figure~\ref{fig:waterfall_beta_dec_bc_Toff} shows the best fit values of the spectral index as a function of LST and observing day.

Only data above $\nu_{\rm min}=60$~MHz were used in the fit. We refer to appendix~\ref{app:stability} for a discussion on the stability of the results with respect to this choice.
We compute the RMS of the residuals for each LST bin and for each day, and report their values in the lower panels of Figure~\ref{fig:waterfall_beta_dec_bc_Toff}. We check for outliers in both $\beta$ and RMS, and mask the corresponding spectra after a visual check. This procedure discards only few spectra.

Results are  consistent across the observing window, highlighting the good stability of the instrument. The value of the spectral index $\beta$ steepens around LST~$\sim13$~h for both antennas. 
The smallest RMS values (around 20~K) occur in the $9-13$~h LST range, motivating further the choice of this interval for further cosmological analysis. 
Similar trends are found for antenna 252A, although the RMS values are overall higher, reaching almost 100~K when the Galactic center is in the sky.
Note that antenna 254A shows persistent variations at large LST values, in correspondence with high RMS values (around 30~K).

\begin{figure}
\includegraphics[width=\columnwidth]{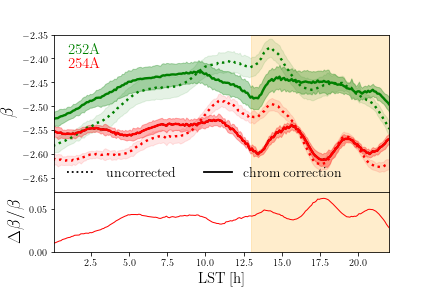}
\includegraphics[width=\columnwidth]{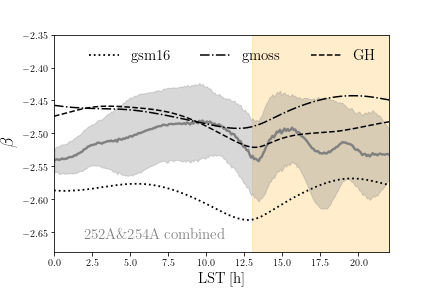}
\caption{{\it Upper panel.} Mean value of the best fit daily spectral index $\beta$, as a function of LST, for the beam corrected reference dataset of Figure~\ref{fig:waterfall_beta_dec_bc_Toff} (solid lines) compared with the non-corrected one (dotted lines), for the antenna 254A (in red) and 252A (in green). Shaded areas show the standard deviation. The bottom part of the Figure show the relative difference between the result from antenna 252A and 254A. 
{\it Lower panel.} The combined result of the two antennas (solid gray line) and its scatter (shaded area) is compared with  
the Guzman-Haslam (black dashed), improved GSM (black dotted) and GMOSS (black dashed dotted) spectral index (see text for details). For illustrative purposes, the shaded yellow area indicates 
daylight, starting approximately from sunrise.}
\label{fig:beta_Toff}
\end{figure}

We show in the upper panel of Figure~\ref{fig:beta_Toff} the value of the mean spectral index as a function
of LST and its variance across the different days of the dataset. 
The two antennas show similar trends although their results are slightly offset, with antenna 252A pointing toward a flatter $\beta$, as visible also from the upper panels of Figure~\ref{fig:waterfall_beta_dec_bc_Toff}.
The mean spectral index shows almost no dependence on LST in the range $0-10$~h for antenna 254A and in the range $5-10$~h for antenna 252A.  
It then becomes steeper for both antennas around LST~$\sim13$~h.
Antenna 252A shows an almost constant spectral index for LST$>15$~h, while we can recognize the more complex pattern discusthat sed above for antenna 254A.
The bottom part of the Figure shows that the relative variation between the two antennas that is always lower than 6\%.
We report in Table~\ref{tab:ref_beta} the mean value of the spectral index and its standard deviation averaged in LST bins of $3$~h.
We also  report the results obtained without the beam factor correction,  to assess its impact.
The spectral index in this case is found to flatten when the sky temperature is minimal, in agreement with the results in \citet{Price2018}, for which the beam chromaticity was not taken into account. The effect of the correction is stronger  below LST$\sim 5$~h and around LST~$\sim13$~h. 

The quoted spectral index scatter in Table~\ref{tab:ref_beta} and upper panel of Figure~\ref{fig:beta_Toff} is derived directly from the standard deviation of the single antenna measured spectra. The data scatter alone is not enough to reconcile the discrepancy between antenna 254A and 252A, possibly due the presence of a (yet) unknown systematic effect. To take this into account we consider together the data from the two antennas. The combined result is shown
in the lower panel of Figure~\ref{fig:beta_Toff}. 
Measurements from the two separate antennas have similar trends in LST, implying that the previous considerations remain valid for the combined results. The mean spectral index $\beta$, combining the steeper results from antenna 254A and the flatter results from antenna 252A, varies between $-2.54$ and $-2.48$. The uncertainty, now computed the from the standard deviation of all the spectra, has a magnitude of $\Delta \beta \sim 0.06$ across the full LST range, reaching $\Delta \beta \sim 0.08$ when the Galactic center is above the horizon.
We compare this result with existing measurements and simulations.
Following \citet{Mozdzen2019}, we first compute a spectral index
from the comparison of the Haslam and the Guzman maps:
\begin{equation}\label{eq:betaGH}
    \beta_{\rm GH}({\rm LST})=\ln{\frac{T^{\rm ant}(\nu_{45},{\rm LST})}{T^{\rm ant}(\nu_{408},{\rm LST})}} \, \left( \ln{\frac{45}{408}} \right)^{-1},
\end{equation}
where $T^{\rm ant}(\nu, {\rm LST})$ is the portion of the sky model seen though the beam as the sky drifts, and thus is  a function of LST.
Moreover, we consider both the improved version of the GSM \citep{Zheng2017} and the GMOSS \citep{GMOSS2017} to create a set of full sky maps in the $60-87$~MHz range, with 1~MHz spacing. We then simulate an observation with our antenna beam and consider LST values 10~minutes apart.  We construct with this procedure a mock spectrum $T^{\rm ant}(\nu, {\rm LST})$ for every LST value. Note that we consider the beam shape at the reference frequency of $75$~MHz to avoid introducing chromaticity effects. We then perform a non-linear least squares minimization to constrain the model parameters of equation~\ref{eq:model_beta} for each LST bin, and report the best-fit spectral index in  Figure~\ref{fig:beta_Toff}.

The overall variation with LST of these models and of $ \beta_{\rm GH}({\rm LST})$ is in good agreement with the one found in our data. Note for example the steeper spectral index around LST~$\sim13$~h. 
It is worth noticing that the best-fit values for the scaling temperature $T_{75}$ are well in agreement with their observed ones, once corrected for the offset. This assures that the best-fit solution is not falling in un-physical minima.

\begin{table}
\caption{Spectral index $\beta$ (equation~\ref{eq:model_beta}) averaged in $3$~h LST bins from the {\it reference} dataset. Errors $\Delta \beta$ are computed from the standard deviation of all the spectra. The RMS is also the mean of the RMS values within each bin.}\label{tab:ref_beta}
\begin{center}
\begin{tabular}{|c|c|c|c|c|c|c|} 
\hline
  & \multicolumn{3}{|c|}{254 A} & \multicolumn{3}{|c|}{252 A} \\
\hline
LST (h) & $\beta$ & $\Delta \beta$ & RMS (K) & $\beta$ & $\Delta \beta$ & RMS (K) \\
 \hline
0-3 & -2.55 & 0.01 & 25.08 & -2.51 & 0.02 & 112.37 \\
3-6 & -2.55 & 0.01 & 21.99 & -2.46 & 0.02 & 94.14 \\
6-9 & -2.54 & 0.01 & 21.90 & -2.43 & 0.02 & 82.58 \\
9-12 & -2.54 & 0.01 & 20.70 & -2.44 & 0.02 & 79.05 \\
12-15 &-2.58 & 0.01 & 23.31 & -2.46 & 0.02 & 93.77 \\
15-18 & -2.57 & 0.01 & 29.33 & -2.45 & 0.02 & 135.40 \\
18-21 & -2.58 & 0.01 & 26.96 & -2.46 & 0.02 & 153.21 \\
\hline
\end{tabular}

\end{center}
\end{table}

\subsection{All dataset results}\label{sec:beta_6datasets}

We now analyze the value of the spectra index for the spectra in the 6 {\it datasets} defined in Section~\ref{sec:obs}, following the same procedure of Section~\ref{sec:beta_reference}. Although the data show a coherent behavior on a time scale of few tens of days, there are changes throughout the months that reflect in different $\bar{T}_{\rm off}$ values for each dataset.
Slight variations across {\it datasets} are thus expected for the best-fit values of the spectral index. 

\begin{figure}
\includegraphics[width=\columnwidth]{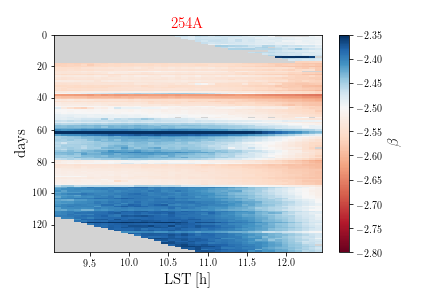}\\
\includegraphics[width=\columnwidth]{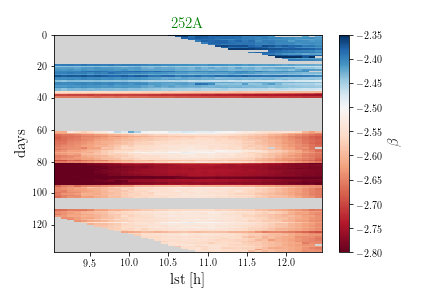}
\caption{Best fit value for the $\beta$ parameter of equation~\ref{eq:model_beta} for every day of our full dataset in the LST range 9-12.5 h (LST binning $\sim5$~mins) for the antenna 254A (upper panel) and 252A (lower panel). Note that data from May 2018 and 2019 do not cover the entire LST range, to avoid data too close to the sunset. 
The colorbar is the same for the two antennas to aid the comparison.}
\label{fig:waterfall_beta_all}
\end{figure}

We report these values of $\beta$ as a function of LST and observing day in Figure~\ref{fig:waterfall_beta_all}, and in Table~\ref{tab:datasets_beta} the mean and standard deviation for each dataset. We also report the averaged RMS values of the residuals. 
Note that, in the restricted LST range $9-12.5$~h, we can enlarge the {\it reference} dataset with more spectra that we are otherwise obliged to discard when requiring high quality over the full $24$~h range. For this reason dset2 has larger error bars than the one presented in Section~\ref{sec:beta_reference}.

\begin{figure*}
\includegraphics[width=\columnwidth]{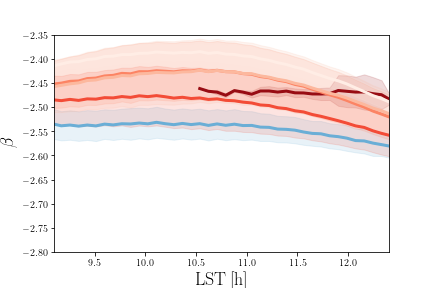}
\includegraphics[width=\columnwidth]{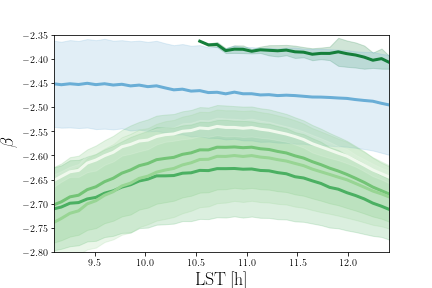}\\
\includegraphics[width=\columnwidth]{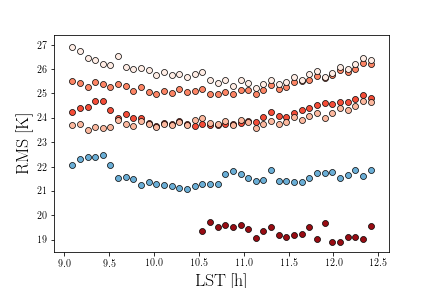}
\includegraphics[width=\columnwidth]{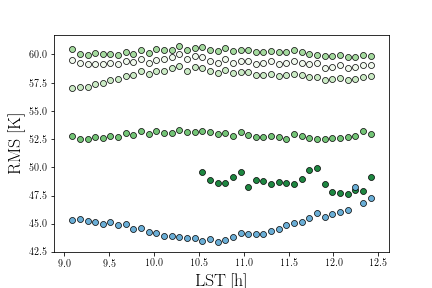}

\caption{{\it Upper panel:} Mean and standard deviation of the best fit values of $\beta$ of Figure~\ref{fig:waterfall_beta_all} split between the different datasets defined in Section~\ref{sec:obs}, color coded as in Figure~\ref{fig:DT_nu}, for antenna 254A (left, red scale) and antenna 252 (right, green scale). {\it Lower panel:} Mean RMS for the residuals for antenna 254A (left) and 252A  (right). Every marker represents a different LST bin ($\sim5$ min large) color coded to identify the different datasets. The {\it reference} dataset is reported in blue to facilitate identification.}
\label{fig:beta_RMS_split}
\end{figure*}

We report in Figure~\ref{fig:beta_RMS_split} the mean and standard deviation obtained from Figure~\ref{fig:waterfall_beta_all} for each dataset, as a function of LST. There is good consistency across datasets and we find 
 similar variation as a function of LST, in particular for antenna 254A.
We also report in the lower panels the variation of the RMS of the residuals. The RMS values are always lower for antenna 254A than for antenna 252A, as found for the case of the {\it reference} dataset. We do not report the best-fit value of $T_{75}$ but we remark its agreement with the (offset corrected) measured value of the sky temperature at 75 MHz.

\begin{figure}
\includegraphics[width=\columnwidth]{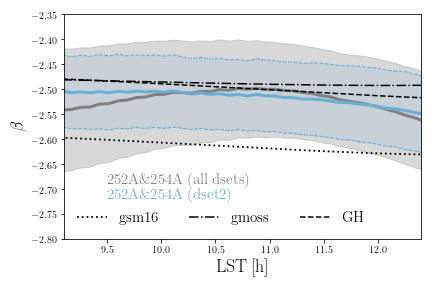}
\caption{Combined result from all {\it datasets} and antennas (solid gray line) and its scatter (shaded area) compared with  
the Guzman-Haslam (black dashed), improved GSM (black dotted) and GMOSS (black dashed dotted) spectral index (see Section~\ref{sec:beta_reference} for details). The combined results for dset2 alone is also shown (solid blue line) together with its standard deviation (shaded blue area).}
\label{fig:dsetcombined}
\end{figure}

Similarly to what done in Section~\ref{sec:beta_reference}, we show in Figure~\ref{fig:dsetcombined} the combined spectral index $\beta$ and its standard deviation considering both antennas and all {\it datasets}. The mean value of $\beta$ varies from a minimum of $-2.50$ at LST~$\sim 11$~h to a steeper $-2.56$ at LST~$\sim 12.5$~h. The combined error bar spans the $0.09 < \Delta \beta < 0.12$ range. We compare the result with existing measurements and simulations.

 \begin{table}
\caption{Mean of the best-fit spectral index $\beta$ (equation~\ref{eq:model_beta}) for each of the six {\it dataset}s, obtained by  averaging over the LST range $9-12.5$~h. The quoted RMS values are averaged accordingly. The errors $\Delta \beta$ are computed from the standard deviation of the spectra.  }\label{tab:datasets_beta}
\begin{center}
\begin{tabular}{|c|c|c|c|c|c|c|} 
\hline
  & \multicolumn{3}{|c|}{254 A} & \multicolumn{3}{|c|}{252 A} \\
\hline
{\it dataset} & $\beta$ & $\Delta \beta$ & RMS (K) & $\beta$ & $\Delta \beta$ & RMS (K) \\
 \hline
1 & -2.47 & 0.01 & 19.32 & -2.39 & 0.01 & 48.72 \\
2 ({\emph ref}) & -2.54 & 0.03 & 21.60 & -2.47 & 0.10 & 44.88 \\
3 & -2.50 & 0.06 & 24.15 & -2.66 & 0.09 & 52.86 \\
4 & -2.44 & 0.06 & 25.35 & -2.63 & 0.08 & 60.20 \\
5  & -2.45 & 0.06 & 23.90 & -2.65 & 0.09 & 58.11 \\
6 & -2.41 & 0.02 & 25.88 & -2.58 & 0.02 & 59.30 \\
\hline
\end{tabular}

\end{center}
\end{table}

\section{Discussion and conclusions}
\label{sec:conclusions}

We presented the analysis of LEDA radiometric data with the aim of measuring the spectral index of Galactic radio emission in the $60-87$~MHz range. Observations were carried out with two antennas (252A and 254A) and spanned a total of $\sim 600$ hours divided in six different {\it datasets} covering May 2018 and from mid-December 2018 to mid-May 2019.

Observed spectra were corrected for the effect of beam chromaticity following \citet{Mozdzen2017,Mozdzen2019} and re-scaled to be in agreement with the absolute temperature scale derived from existing observations of \citet{Guzman2011,Haslam1982}. The beam chromaticity effect leads to spectral index corrections of no more than $\sim 4\%$. We note here that an approach that incorporates beam effects in the foreground model and, therefore, bypasses the chromaticity correction, has recently been discussed in the literature \citep{Tauscher2020,Anstey2020}, although not applied to observations yet.   

Due to optimal soil conditions, we first focused on 14 days of data from December 2018 and early January 2019 ({\it reference} dataset). We found that the spectral index is fairly flat across 24~h LST range. It varies from $\beta = -2.55 \pm 0.01$ for antenna 254A at LST~$< 6$~h to $\beta = -2.58 \pm 0.01$ at LST~$\sim 13$~h.  
A similar behavior is found for antenna 252A, although with a flatter mean spectral index, changing from $\beta = -2.43 \pm 0.02$ to $\beta = -2.46 \pm 0.02$. Residual rms values after subtracting the best fit power law vary from $20-30$~K for antenna 254A to $70-150$~K for antenna 252A. 

Results from the two antennas show a $\sim 5 \%$ offset which is larger than their individual uncertainties.
We thus jointly analyze measurements from the two antenna to account for this systematic offset.
The combination of the two antennas leads to values of the spectral index ranging from $\beta=-2.48$ at LST~$\sim 10$~h to a steeper $\beta=-2.54$ at LST~$\sim 13$~h, with a typical standard deviation of $\Delta \beta \sim 0.06$.\\

Our results are broadly consistent with previous measurements. \citet{Guzman2011} derived a spectral index map from 45~MHz and 408~MHz with $-2.6 < \beta < -2.5$ over most of the sky.
Similarly, \citet{Mozdzen2019} found $-2.59 < \beta < -2.54$ at LSTs from $0$ to $12$~h in the $50-100$~MHz range, flattening to $\beta = -2.46$ when the Galactic center is transiting - although their observations cover the southern celestial hemisphere and can, therefore, only be qualitatively  compared to ours.
Moreover, we find broad agreement with the Improved GSM \citep{Zheng2017} and GMOSS \citep{GMOSS2017} sky models. 

We analyzed the full six {\it dataset}s in the $9-12.5$~h LST range, where the foreground temperature shows a  minimum.
The distribution of the best fit spectral indexes for antenna 254A spans values from $-2.54$ to $-2.41$, all consistent with the mean $\beta = -2.48 \pm 0.07$. Spectral indexes are in general less stable for antenna 252A, leading to a mean value $\beta = -2.6 \pm 0.1$. \\
The configuration of antenna 252A was not the same throughout the observing period and a larger scatter is thus expected in its results. Improved antenna simulations, including a more refined modeling for the ground plane and the soil,
are ongoing and will be the subject of a follow up work.\\

We combined together the six {\it datasets} from both antennas and find $\beta = -2.5 \pm 0.1$ in the $9-12.5$~h LST range. The combined mean spectral index value for all our data is in agreement with the result obtained using the {\it reference} dataset (dset2) and consistent with existing sky models.

%%%%%%%%%%%%%%%%%%%%%%%%%%%%%%%%%%%%%%%%%%%%%%%%%%%%%%%%%%%%%%%%%%%%%%%%%%%%%%%%

\section*{acknowledgements}
We thank the anonymous referee for their useful and constructive suggestions.
The authors kindly thank the Caltech OVRO staff for the great dedication and skills demonstrated in constructing the LWA array.
MS acknowledges funding from the INAF PRIN-SKA 2017 project 1.05.01.88.04 (FORECaST) and support from the INFN INDARK PD51 grant. AF is supported by the Royal Society University Research Fellowship.
LEDA research has been supported in part by NSF grants AST/1106059, PHY/0835713, and OIA/1125087. The OVRO-LWA project was enabled by the kind donation of Deborah Castleman and Harold Rosen.
This research made use of \texttt{Numpy} \citep{Numpy2020}, \texttt{Astropy} \citep{Astropy} and \texttt{Scipy} \citep{Scipy2020}.
Some of the results in this paper have been derived using the \texttt{healpy} \citep{Zonca2019} and \textit{HEALPix} \citep{2005ApJ...622..759G} package.

\section*{Data availability}

The data underlying this article will be shared on reasonable request to the corresponding author.

\bibliographystyle{mnras}
\bibliography{biblio}

%%%%%%%%%%%%%%%%% APPENDICES %%%%%%%%%%%%%%%%%%%%%

\appendix

\section{Observation details}\label{app:data}
We detail in this appendix the total amount of hours available for our analysis. We recall that the observing season spanned from December 2018 to May 2019. Data from May 2018 were also included. The final dataset consists of 137 days of usable data. For each day, not all LST are considered. We mainly focus on the $9-12.5$ range as depicted in Figure~\ref{fig:sun}. We expand for 14 selected days in December/January the LST range to an almost 24~h LST range. We report the total amount of hours in Table~\ref{tab:tot_h} dividing the LST range  according to this.

\begin{figure}
\includegraphics[width=\columnwidth]{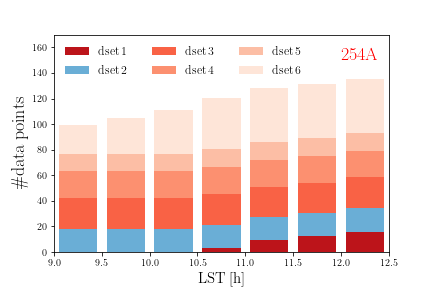}\\
\includegraphics[width=\columnwidth]{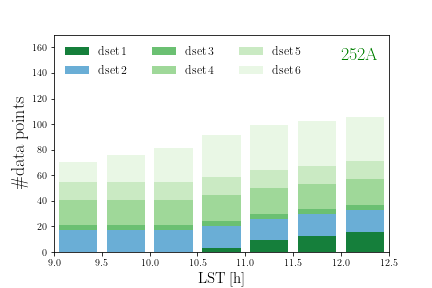}
\caption{Total number of data points for each frequency in LST slots of $\sim 0.5$~h in the preferred $9-12.5$~h range, for the antenna 254A (top panel) and the antenna 252A (bottom panel). We highlight how they are distributed between the different \textit{datasets} following the same color code of Figure~\ref{fig:DT_nu}.}
\label{fig:tot_h}
\end{figure}

As discussed in Section~\ref{sec:obs}, the 137 days are split in 6 different \textit{datasets}. We detail the number of observing days for each of them in Table~\ref{tab:dsets} and the total amount of data available per LST slot in Figure~\ref{fig:tot_h}.
Note that the {\it reference} dataset contains only a subset of the spectra of {\it dset2}, as few spectra were discarded in the analysis of the larger LST range.
\begin{table}
\caption{Total hours of data analyzed for the antenna 254A and 252A divided in different LST slots.}\label{tab:tot_h}
\begin{center}
\begin{tabular}{|c|c|c|} 
\hline
LST (h) & 254A (hours of data) & 252A (hours of data) \\
\hline
0h-9h & 126 & 126 \\
9h-12h & 411 & 321 \\
12h-24h & 140 & 140 \\
\hline
Total & 677  & 587 \\
\hline
\end{tabular}

\end{center}
\end{table}

\begin{table}
\caption{Number of days observed in each data set. Note that each data set can extend beyond a calendar month.}\label{tab:dsets}
%\caption{Number of days observed in each campaign. The campaign can extend beyond a calendar month. for each month of data taking and the observing period composing the six \textit{datasets} used in this work}\label{tab:dsets}
\begin{center}
\begin{tabular}{|c|c|c|c|} 
\hline
dset & month(s) & 254A: days & 252A: days \\
\hline
 1 & May2018 & 17 & 17 \\ 
 2 (ref) & Dec2018/Jan2019 & 19 (14) & 17 (14) \\
 3 & Jan/Feb2019 & 24 & 12 \\
 4 & Feb/Mar2019 & 21 & 20 \\
 5 & Mar2019 & 14 & 14 \\
 6 & Mar/Apr/May2019 & 42 & 35 \\
\hline
\end{tabular}

\end{center}
\end{table}

%%%%%%%%%%%%%%%%%%%%%%%%%%%%%%%
%%%%%%%%%%%%%%%%%%%%%%%%%%%%%%%%%%%%%%%%%%%%

\section{Stability of the results on the spectral index}\label{app:stability}
We perform some further tests to assess the stability of our measured spectral indices. We concentrate on the results of Section~\ref{sec:beta_reference} which are the central ones of this work.

\begin{figure}
\includegraphics[width=\columnwidth]{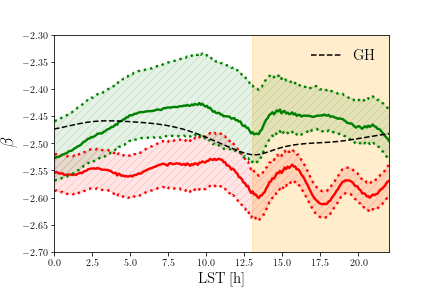}
\caption{The mean value of the beam corrected best fit daily spectral index of Figure~\ref{fig:beta_Toff} (254A in red and 252A in green) compared with 
the best fit obtained from equation~\ref{eq:chi} substituting $T_{\rm off}$ with  $T_{\rm off}+1\sigma$ and $T_{\rm off}-1\sigma$ from Figure~\ref{fig:meanToff} (dotted lines). The dashed area shows the range of values of $\beta$
from intermediate values of $T_{\rm off}$.
For comparison the Guzman-Haslam spectral index is shown (black dashed line, see equation~\ref{eq:betaGH}). For illustrative purposes, the shaded yellow area indicates daylight, starting approximately from sunrise.}
\label{fig:beta_Toff_ul}
\end{figure}

\smallskip First we investigate the dependence of our findings with respect to the offset $T_{\rm off}$ discussed in Section~\ref{sec:offset} and reported in Figure~\ref{fig:meanToff}. There is an expected correlation between the offset and the steepness of $\beta$: the larger the offset that is removed, the more we should obtain a steep $\beta$, since the subtraction is more effective at high frequencies, where the sky temperature is lower. We report results obtained considering the 1$\sigma$ deviation from the mean $T_{\rm off}$ in Figure~\ref{fig:beta_Toff_ul}. The result for every intermediate value falls in between these solutions for both antenna 254A and 252A. The reported variations shows  a consistent trend with LST and deviations of no more than a few \% for the overall value of the spectral index.

\begin{figure}
\includegraphics[width=\columnwidth]{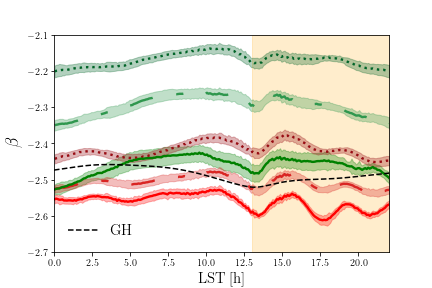}
\caption{The mean value of the beam corrected best fit daily spectral index of Figure~\ref{fig:beta_Toff} (solid red line for antenna 254A and solid green for the 252A) compared with 
the best fit obtained by  varying the minimum frequency considered in the fit, from the standard $60$~MHz to $55$~MHz (dot-dashed lines) and $50$~MHz (dotted lines). 
The red and green shaded areas show the standard deviation.
For comparison the Guzman-Haslam spectral index is shown (black dashed line, see equation~\ref{eq:betaGH}). For illustrative purposes, the shaded yellow area indicates daylight, starting approximately from sunrise.}
\label{fig:beta_fmin}
\end{figure}

\smallskip Our results for the spectral index have been obtained considering a minimum frequency cut  at $\nu_{\min}=60$~MHz for our spectra. This choice, already discussed in Section~\ref{sec:obs}, minimizes the environmental effects in the data. 
We report in Figure~\ref{fig:beta_fmin} the best fit of the spectral index for the {\it reference} dataset obtained by  lowering $\nu_{\min}$ to $55$ or $50$~MHz, for completeness.
We find a flattening of the spectral index, expected when including these lower frequencies \citep[e.g.][]{deOliveira2008}
The effect is particularly strong for antenna 252A, reaching $\beta\sim -2.15$. Note however that the trend in LST is consistent for all frequency cuts. The results for antenna 254A are more stable, as also expected from Figure~\ref{fig:DT_nu}.

\bsp	% typesetting comment
\label{lastpage}
\end{document}